# A TLM-BASED PLATFORM TO SPECIFY AND VERIFY COMPONENT-BASED REAL-TIME SYSTEMS

Mostafavi Amjad Davoud and Zolfy Lighvan Mina

Department of Electrical and Computer Engineering, Tabriz University, Tabriz, Iran

## ABSTRACT

*This paper is about modeling and verification languages with their pros and cons. Modeling is dynamic part of system development process before realization. The cost and risky situations obligate designer to model system before production and modeling gives designer more flexible and dynamic image of realized system. Formal languages and modeling methods are the ways to model and verify systems but they have their own difficulties in specifying systems. Some of them are very precise but hard to specify complex systems like TRIO, and others do not support object oriented design and hardware/software co-design in real-time systems. In this paper we are going to introduce systemC and the more abstracted method called TLM 2.0 that solved all mentioned problems.*

## 1. INTRODUCTION

Using existing components to assemble new software systems that called component based software development (CBSD) can effectively improve software development efficiency, reduce development costs and improve software's quality and reliability [1]. Furthermore, by applying some constraints such as time we can specify real-time system too. By CBSD method designer could make the real-time system by using reliable, high-quality and low cost COTS very rapidly just by connecting them together and test it as a whole system. This approach is an abstract way to solve problems and takes you away from details of system that are going to distract you from the main goal.

The paper is organized as follows. Section 2 provides an overview of the related work and includes the details of the selected languages. Section 3 introduces SystemC and TLM 2.0 method and case study. In Section 4 languages are evaluated and summarize the pros and cons of languages. Section 5 includes the conclusion of the paper.

## 2. RELATED WORKS

There are a lot of languages and modeling methods to design and verify real-time systems. Each of them has its own characteristics. In coming sections we are going to discuss them.

## 2.1. FORMAL LANGUAGES

The history of formal languages is very old. A lot of formal languages proposed in the past and todays, designer are using them to specify systems. Formal representation increases the confidence on the system in early stages of development. Especially in the context of Real-Time systems which are mostly safety critical systems, the correctness of specification is very important.

Different formal methods have been proposed for specifying and verifying timed systems such as state-based, model-based, logic based and net-based [2].

**2.1.1. TRIO** is a logic based specification language. It is an extension to classical first order temporal logic. The disadvantage of TRIO is that it is hard to understand because of its complicated syntax that is close to machine level language [2].

**2.1.2. Real time Object Z** is a combination of the formal specification language Object Z and timed calculus. The timed calculus allows you to specify and manipulate time but it does not support standard object oriented design [2].

**2.1.3. Timed Automata** is another formalism that is evolving now a day for timed systems. Timed Automata provides good structures to specify time. however it is difficult to model complex systems as an automaton. Also a slight change in specification may lead to a huge change in model [2].

**2.1.4. VDM++** (Vienna Development Method) is a formal specification language intended to specify object oriented systems with parallel and real-time behavior, typically in technical environments. The language is based on VDM-SL, and has been extended with class and object concepts, which are also present in languages like Smalltalk-80 and Java. This combination lets developers to use object oriented formal specifications in there design approach [3].

## 2.2. MODELING

We are going to introduce some modeling methods to specify real-time systems. All of them can specify system but they can not specify software parts of the designed system. Then the designer and developer should work on two separated system as hardware and software parts.

**2.2.1 MBI&T Method**, The model-based integration and testing method or MBI&T method is shown in Figure 1. Depending on the component representations that are integrated, i.e., only models, combined models and realizations, or only realizations, a different type of infrastructure $I$ may be required. When only models are integrated, the infrastructure model $I_M$ is used and when combined models and realizations are integrated, a model-based integration infrastructure $I_{MZ}$ is used [4].

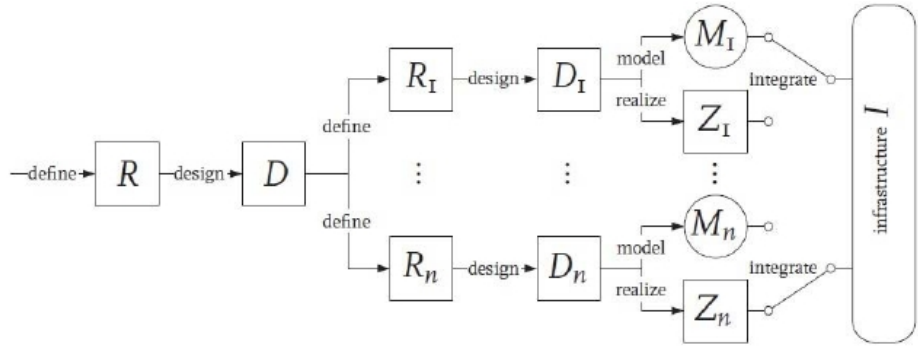

Figure 1. System development process in the MBI&T method [4].

**2.2.2. DARTS**, a design approach for real-time systems, ADARTS (DARTS extension for Ada-based systems), and COMET (DARTS extension to UML). They decompose real-time systems into tasks. In DARTS and its variants, a real-time system is first decomposed into tasks then they group into software modules. DARTS, therefore, helps the designer to make the transition from tasks to modules. However, it does not provide mechanisms for checking and verifying temporal behavior of a system under development [5].

**2.2.3. TRSD**, a transactional real-time system design, is an approach to real-time system development that introduced by Kopetz et al. Real-time systems Building blocks are transactions that consist of one or more tasks. They are related to real-time attributes, e.g., deadline and criticality. TRSD, has rules for decomposition of real-time systems into tasks, and provides temporal analysis of the overall real-time system [5].

# 3. PROPOSED APPROACH

**3.1. SystemC** is a system design language that improves overall productivity for designers of electronic systems. Today's systems contain application-specific hardware and software parts. Furthermore, the hardware and software are usually co-developed on a tight schedule, the systems have tight real-time performance constraints, and thorough functional verification is required to avoid expensive and sometimes catastrophic failures [6].

SystemC offers real productivity gains by letting engineers design both the hardware and software components together at a high level of abstraction. This higher level of abstraction gives the design team a complete understanding of system early in the design process of the entire system and enables better system tradeoffs, better and earlier verification, and overall productivity gains through reuse of early system models as executable specifications [6].

Strictly speaking, SystemC is a class library within a well-established language, C++. However, when SystemC is coupled with the SystemC Verification Library, it does provide in one language many of the characteristics relevant to system design and modeling tasks [6].

The following diagram illustrates the major components of SystemC [6].

| User libraries | SCV | | Other IP |
|---|---|---|---|
| Predefined Primitive Channels: Mutexs, FIFOs, and Signals | | | |
| Simulation Kernel | Threads & Methods | Channels & Interfaces | Data types: Logics, |
| | Events, Sensitivity & Notifications | Modules & Hierarchy | Integers, Fixed points |
| **C++** | | | **STL** |

Figure 2. SystemC language Architecture [6].

**3.2. TLM 2.0** is the transaction level modeling approach. And the main question that ESL struggling with is, "what is the most appropriate taxonomy of abstraction levels for transaction level modeling?" In real projects, models have been categorized according to a range of criteria, including granularity of time, frequency of model evaluation, functional abstraction, communication abstraction, and use cases. The TLM-2.0 activity explicitly recognizes the existence of a variety of use cases for transaction-level modeling but rather than defining an abstraction level around each use case, TLM-2.0 takes the approach of distinguishing between interfaces (APIs) on the one hand, and coding styles on the other [7].

TLM-2.0 consists of a set of core interfaces, the global quantum, initiator and target sockets, the generic payload and base protocol, and the utilities. The TLM-1 core interfaces, analysis interface and analysis ports are also included. Its core interfaces consist of the blocking and non-blocking transport interfaces, the direct memory interface (DMI), and the debug transport interface. The generic payload supports the abstract modeling of memory-mapped buses, in addition to them the extension mechanism support the modeling of specific bus protocols that maximizes interoperability [7].

The TLM-2.0 classes are more abstracted layer on top of the SystemC class library [7]. As you see in Figure- 3. From up to down everything get more sophisticated and concentrates on details. It causes designers make more mistakes during modeling of systems. But from bottom up of pyramid everything gets more abstracted and modeling of system being easier.

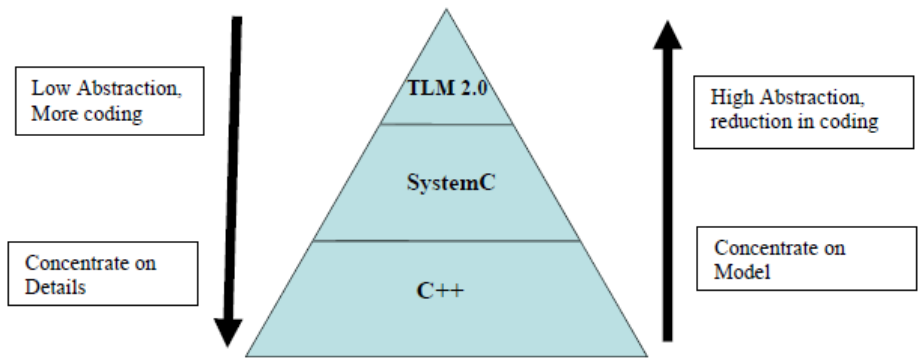

Figure 3. The pyramid of abstraction benefits

For maximum interoperability, and particularly for memory-mapped bus modeling, using the TLM-2.0 core interfaces, sockets, generic payload and base protocol together is recommended. These classes are known collectively as the interoperability layer. In some cases that using generic payload is not appropriate, it is possible to use core interfaces and the initiator and target sockets, or the core interfaces alone. They can be used with an alternative transaction type. It is possible to use the generic payload without initiators and target sockets and just by using core interfaces, although this approach is not recommended [7].

Initiators can start sending transaction object with read/write command toward targets and targets do the commands and send back the modified object toward senders. Figure 4 and 5 show the sample architectures that can be used in specifying systems.

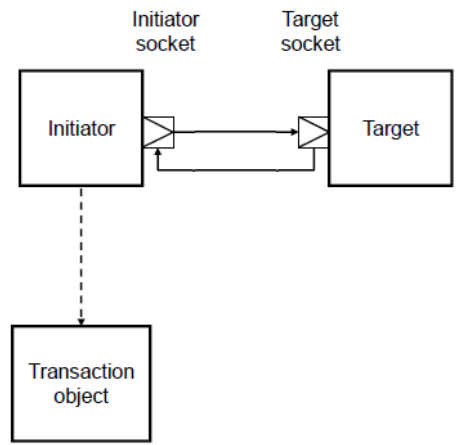

Figure 4. A sample TLM 2.0 structure

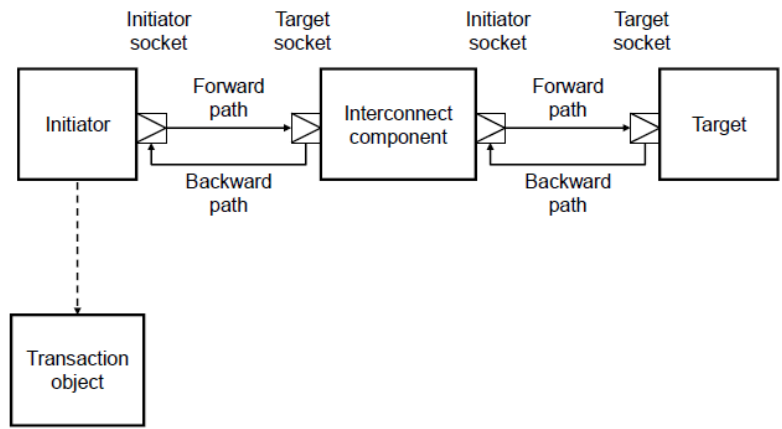

Figure 5. A sample TLM 2.0 structure [7].

The definition of the standard TLM-2.0 interfaces and coding style are different. The standard interfaces of TLM-2.0 ensure the interoperability of system [7].

### 3.2.1. Advantages and disadvantages of TLM based modelling

Advantages:

- The TLM-2.0 classes are layered on top of the SystemC class library [5]. The designer can specify complex systems that are a hybrid between hardware and software [8].

- The TLM interface standard enables SystemC model communication and reuse at the transaction level. It provides an essential ESL framework for architecture analysis, software development, software performance analysis, and hardware verification [9].

- With TLM method you can make software and hardware parts of expected system simultaneously.

- The TLM standard based on C++ programming language that has a standard syntax.

- Because of using C++ language in TLM. Developers can use other APIs of C++ to connect models to outside world. Then they can design a system with parts model and other parts realized one.

- It is object oriented and we can wrap classes into each other to change behavior of modules as components.

- With TLM you can model system at higher abstraction level or you can go down tile register transfer level.

Disadvantages:

- The lack of user friendly environment for the TLM is a great challenge in modeling world. This is the issue we have resolved it by developing a user friendly environment.

### 3.2.2. Proposed Platform

We have developed an environment to facilitate modeling in TLM 2.0. In addition, it has some components that are not part of TLM standard but they are simplifying design approach. We have added Cpu's that can modify the timing of every assigned modules to them by scale of their frequency. Furthermore, virtual buses in the environment give the designer a real picture of system under development. They are virtual and just connecting processors to each other.

After construction of cpu's and their connections you have to design the modules and instance of them as components and then assign them to Cpu's. The proposed platform lets you to add 3 types of modules. The initiator modules that have delay, number of sockets, bandwidth, and transaction object. The target modules have delay for each socket because we assume that every target has different computing parts that connected to different sockets, sockets, and bandwidth. The router or interconnector modules have delay, two types of sockets; they act as receiver and sender, and bandwidth. When you make a router you have to design its internal connection too. The developed environment lets you connect sockets as 1 to N approach. Finally you can connect all of the components together and make the system. It generates TLM 2.0 standard codes for your system. After execution of system, it makes log file that can be analysed by platform and shows you the start and end time of every instance of modules as a timing diagram. The designer of system could verify his/her own system timing just by looking at this diagram.

The developed environment lets you add, delete, and modify every module then you can gradually define your system and make it better. At same time you can generate codes and test their result step by step.

### 3.2.2.1. Case Study

We have designed a hypothetical ABS system for automobile in this platform. The components specifications are listed in tables below. Because of simplicity we used simple numbers.

Table 1. Modules name and their characteristics

| Module type | Module name | Delay | Instance of module | Cpu |
|---|---|---|---|---|
| Initiator | Module0 | 10 ns | Brake | Cpu0 |
| Router | Module1 | 5 ns | Router | Cpu1 |
| Target | Module2 | 20 ns | ABSbrake1,2,3,4 | Cpu2,3,4,5 |

Table 2. Cpu's and their ferequency

| Cpu name | Frequency(GHz) |
|---|---|
| Cpu0 | 1 GHz |
| Cpu1 | 5 GHz |
| Cpu2,3,4,5 | 4 GHz |

First, the structure of system has to be created after specifying Cpu's and connecting them at bus section together. Figure 6 shows the system schema that was created by the platform.

Next, Modules are created and codes generated. The codes can be executed in another TLM 2.0 based project. The system logs its timing that can be seen in diagram which our platform generated from file. Figure 7 shows the generated diagram.

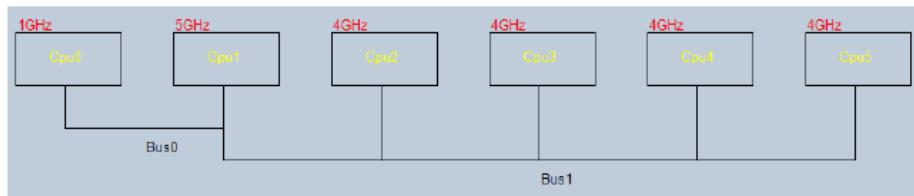

Figure 6. The architecture of system that generated by proposed platform.

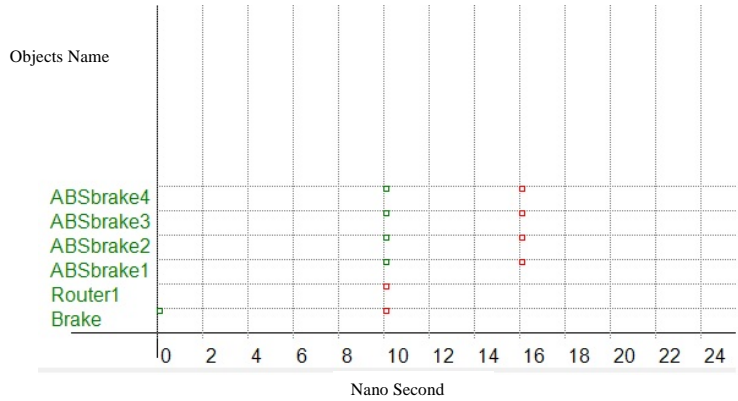

Figure 7. Timing diagram of ABS system.

The green points show start time of every instance and the red one show the end time in Nano seconds. Then we can look at this simple diagram and can see that whole system work at 16 Nano seconds. It shows the system works at expected time that depends on Cpu's frequency and modules internal delays.

## 4. EVALUATION OF LANGUAGES AND MODELS

All of languages and models that mentioned in related work section are good and precise. Each of them has its own terminology to specify the system. When we are going to specify a small system we can use all of them with their difficulties but in large system we have to choose the most abstract system that can specify both software and hardware parts of the system.

Table 3. General evaluation of specification languages [2].

| Language | Type | Handle Concurrency | Synchronization | Executable | Domain |
|---|---|---|---|---|---|
| ASTRAL | State Based | YES, Processes execute Concurrently | NO, Processes ONLY execute asynchronously | YES | Real Time System |
| LUSTER | Dataflow, Descriptive, Programming language | YES | YES | YES | Reactive and synchronous System, Control Devices |
| VDM++ | Object Oriented | YES, concurrent threads are used | YES | YES | Real-Time systems, Control Systems |
| Z | State Based | NO | No Built In support | NO | Not specific |
| Object Z | Object Oriented | YES | YES | NO | Not specific |
| RAISE | Semi Object Oriented | YES | NO | NO | Embedded, safety critical system |
| CSP | Event Based , Process Algebra | YES | YES, Sync between process | NO | Distributed systems |
| Timed Automata | State Based | YES | YES | YES | Not specific |
| SystemC and TLM2.0 | Object Oriented | YES | YES | YES | Not specific |

# 5. CONCLUSIONS

Modeling gives designer a dynamic view of system and reduces costs of system development and their risk in real world. For this purpose a lot of languages and methods have been proposed. Most of them can specify small system very clearly and exactly. However, when systems grow up and get bigger they encounter state explosion problem because of numerous internal states. To overcome this difficulty we have to choose more abstract models and languages to get better picture of real system. All languages except SystemC that we have mentioned above could not specify the actual working software part of real-time system under the development and developers have to design and test their software after realization of the system. Furthermore, in SystemC you can design your system's software in C or C++ language that is common in developing system program and use it in modeled system. In addition you can connect your model to outside world with C++ APIs and implement the mentioned MBI&T method. In addition, TLM 2.0 is more abstract than systemC and has the power of its ancestors. This is power of co-development and testing software and hardware of system in same language.

# REFERENCES

[1]  Yangli Jia, Zhenling Zhang, Shengxian Xie. 2010. Formal Specification and Verification of Components' Real-time Behavior. IEEE International conference on software. 198-201.

[2]  Sammi Rabia, Rubab Iram, Aasim Qureshi Muhammad. 2010. Formal Specification Languages for Real-Time Systems. IEEE. 1642-1647.

[3]  Peter Gorm Larsen, Kenneth Lausdahl, Nick Battle. 2010. VDM-10 Language Manual.

[4]   Cornelis Wilhelmus Niels, Braspenning Maria. 2008. Model-Based Integration and Testing of High-Tech Multi-Disciplinary Systems. PhD thesis. Technische Universiteit Eindhoven.

[5]   Tesanovic and J. Hansson. 2004. Structuring Criteria for the Design of Component-Based Real-Time Systems, in Proceedings of the IADIS International Conference on Applied Computing, Lisbon, Portugal, pp. 1401-1411.

[6]   C. Black David, Donovan Jack. 2004. Systemc from the ground up. Eklectic Ally, Inc. Pp.     263.

[7]   Aynsley John, Doulos. 2009. OSCI TLM-2.0 LANGUAGE REFERENCE MANUAL. Pp 193.

[8]   Approved 10 September 2011. IEEE-SA Standards Board. IEEE Standard for Standard SystemC® Language Reference Manual.

[9]   Accellera website available online at: http://www.accellera.org

**Authors**

**Mostafavi amjad Davoud** received the BS degree from the Tarbiat Moallem University of Tehran (kharazmi now), Tehran, in 2001, and the MS degrees from the University of Tabriz, Tabriz, in 2014, all in computer software engineering.

**Zolfy Lighvan Mina** received the BS degree from the University of Tehran, Tehran, in 1999 in Hardware engineering, the MS from the University of Tehran, Tehran, in 2002 in Computer engineering, and the PHD from University of Tabriz, Tabriz, in 2012 in Electronic engineering (Digital electronic). She is currently faculty of Electrical and Computer Engineering Department, University of Tabriz, Tabriz Iran.